\documentclass{amsart}

\makeatletter
\renewcommand{\email}[2][]{%
	\ifx\emails\@empty\relax\else{\g@addto@macro\emails{,\space}}\fi%
	\@ifnotempty{#1}{\g@addto@macro\emails{\textrm{(#1)}\space}}%
	\g@addto@macro\emails{#2}%
}
\makeatother

\usepackage{graphicx}  % needed for figures
\usepackage{dcolumn}   % needed for some tables
\usepackage{bm}        % for math
\usepackage{amssymb}   % for math
\usepackage{mathrsfs}
\usepackage{graphicx}
\usepackage{natbib}
\usepackage{amsmath}
\usepackage{epstopdf}
\usepackage{color}
\usepackage{soul}
\usepackage{bm}
\usepackage{placeins}
\usepackage{amsmath}
\usepackage{mathtools}
\usepackage{float}

\def\i{{\rm i}}

\usepackage{color}

\def\p{\partial}
\def\be{\begin{eqnarray}}
	\def\ee{\end{eqnarray}}

\def\lp{\left(}
\def\rp{\right)}
\def\lb{\left[}
\def\rb{\right]}
\def\lcb{\left\{}
\def\rcb{\right\}}

\def\lap{\nabla^2}

\def\befi{\begin{figure}}
	\def\eefi{\end{figure}}

\def\bce{\begin{center}}
	\def\ece{\end{center}}

\def\ba#1\ea{\begin{align}#1\end{align}}

\def\bsa#1\esa{\begin{subequations}
		\begin{align}#1\end{align} \end{subequations}}

\newcommand\rvec{\bm{r}}

\newcommand\xvec{\bm{x}}
\newcommand\vvec{\bm{v}}
\newcommand\uvec{\bm{u}}
\newcommand\fvec{\bm{f}}

\newcommand\gammav{\boldsymbol{\gamma}}
\newcommand\omegav{\boldsymbol{\omega}}
\newcommand\Omegav{\boldsymbol{\Omega}}

\graphicspath{ {Figures/} }

\usepackage{amsmath,amssymb,graphicx} 
\usepackage{amsaddr}
\usepackage{mathptmx} 
\usepackage{epstopdf}
\usepackage{natbib}
  \def\ba#1\ea{\begin{align}#1\end{align}}

\DeclareMathAlphabet{\mathpzc}{OT1}{pzc}{m}{it}

\usepackage{mathptmx}
\DeclareMathAlphabet{\mathnew}{OMS}{cmsy}{m}{n}
\usepackage{mathtools} % for \text{}
\usepackage{mathrsfs}
\usepackage{graphicx}
\usepackage{amsmath}
\usepackage{epstopdf}
\usepackage{color}
\usepackage{soul}

\definecolor{darkblue}{RGB}{83,0,93}

\newsavebox{\astrutbox}
\sbox{\astrutbox}{\rule[-5pt]{0pt}{20pt}}

\def\p{\partial}
\def\be{\begin{eqnarray}}
	\def\ee{\end{eqnarray}}
\def\bes{\begin{subeqnarray}}
	\def\ees{\end{subeqnarray}}

\def\lp{\left(}
\def\rp{\right)}
\def\lb{\left[}
\def\rb{\right]}
\def\lcb{\left\{}
\def\rcb{\right\}}

\def\lap{\nabla^2}

\def\befi{\begin{figure}}
	\def\eefi{\end{figure}}

\def\bce{\begin{center}}
	\def\ece{\end{center}}

\def\i{\textrm{i}}

\def\ba#1\ea{\begin{align}#1\end{align}}

\def\bsa#1\esa{\begin{subequations}
		\begin{align}#1\end{align} \end{subequations}}

\usepackage{graphicx}% Include figure files
\usepackage{dcolumn}% Align table columns on decimal point
\usepackage{bm}% bold math
\definecolor{darkblue}{RGB}{83,0,93}

\date{\today}% It is always \today, today,
\numberwithin{equation}{section}
\usepackage[margin=1in]{geometry}

\begin{document}

\title[]{\Large{Flow Characteristics of Chlamydomonas result in \\ Purely Hydrodynamic Scattering}}

\author{Mehdi Mirzakhanloo \ \ \& \ \ Mohammad-Reza Alam}
\address{Department of Mechanical Engineering, University of California, Berkeley, California 94720, USA}
\curraddr{}
\email{reza.alam@berkeley.edu}
\thanks{}
\date{}

\renewcommand{\abstractname}{} 

\begin{abstract}
\normalsize{It has long been believed that swimming eukaryotes feel solid boundaries through direct ciliary contact. Specifically, based on observations of behavior of green alga Chlamydomonas reinhardtii it has been reported that it is their ``flagella [that] prevent the cell body from touching the surface" [Kantsler et al. PNAS, 2013]. Here, via investigation of a model swimmer whose flow field closely resembles that of C. reinhardtii, we show that the scattering from a wall can be purely hydrodynamic and that no mechanical/flagellar force is needed for sensing and escaping the boundary.}
\end{abstract}

\maketitle
%%%%%%%%%%%%%%%%%%%%%%%%%%%%%%%%%%%%%%%%%%%%%%%%%%%%%%%%%%%%%%%%%%%%%%%%%%%%%%

Interaction of swimming microorganisms with solid boundaries is vital to numerous biological processes ranging from fertilization \cite{denissenko2012human} to biofilm formation \cite{durham2012division}. 
While the significance of such interactions have been acknowledged extensively \cite[e.g.][]{berke2008hydrodynamic,li2009accumulation,nash2010run,mino2011enhanced,drescher2011fluid,pimponi2016hydrodynamics,guccione2017diffusivity}, the underlying mechanism is yet a matter of dispute. 
Specifically, there is an unresolved debate over whether it is the \textit{short-range steric} or the \textit{long-range hydrodynamic} that primarily rule microorganisms interactions with solid boundaries \cite{berke2008hydrodynamic,li2009accumulation}.

For microorganisms with rear-mounted flagella (``pusher" type swimmers such as E. Coli bacteria and human spermatozoa), recent studies finally put an end to the debate in support of the hydrodynamic interactions \cite{molaei2014failed,sipos2015hydrodynamic}.
However, for the other major group of microorganisms (``puller" type swimmers, i.e. those with front-mounted flagella such as Chlamydomonas reinhardtii) the primary mechanism of surface scattering has still remained unsettled. 

Few recent theoretical and numerical studies \cite{wu2015amoeboid,wu2016amoeboid,de2016understanding} have shown that specific puller-type swimmers (e.g. deformable swimmers with amoeboid motion) can undergo purely hydrodynamic scattering in a channel (termed as `navigation swimming' \cite{wu2015amoeboid}). Whereas, for the case of Chlamydomonas reinhardtii (widely known as the paradigm of puller-type swimmers), it has been believed that the scattering process is mainly governed by contact/flagellar forces rather than hydrodynamic interactions.
Experiments have shown that C. reinhardtii cells can feel and escape a boundary after getting close enough to the wall  \cite{kantsler2013ciliary}. 
Based on a series of visual observations, it has been claimed \cite{kantsler2013ciliary,lushi2017scattering} that contact forces exerted by flagella to the wall drives the interaction. The observation has been further generalized, suggesting that surface scattering of swimming eukaryotes is primarily steric rather than hydrodynamic \cite{kantsler2013ciliary}.
More recent experimental observations \cite{contino2015microalgae}, nevertheless, do not support this claim: in scattering of C. reinhardtii cells from a curved surface, there exist some cases in which the flagella do not even touch the wall \cite{contino2015microalgae}.

Here, we consider a model microswimmer designed in such a way that its flow field closely resembles that of a C. reinhardtii \cite{jalali2014versatile,jalali2015microswimmer}. Specifically, it induces an oscillatory flow field with anterior, side and posterior vortices in the surrounding fluid. These are characteristics of the flow field generated by the green alga C. reinhardtii \cite{guasto2010oscillatory,drescher2010direct}. Through direct computation, we show that this model swimmer feels and escapes the wall similar to C. reinhardtii, without the need for a physical contact with the wall; hence, the scattering is purely hydrodynamic. 

C. reinhardtii is usually categorized under the ``puller" type swimmers, mainly because it induces the flow field of a contractile dipole in the far field during its effective stroke. However, the flow field induced by the cell in its \textit{close} vicinity, which is of particular importance in the microswimmer-wall interactions, is not just a simple puller- or pusher-type: it is an oscillatory flow field that includes side, anterior, and posterior vortices  (see e.g. \cite{guasto2010oscillatory,drescher2010direct}).

\begin{figure}[h]
	\centering
	\includegraphics[width=0.65\textwidth]{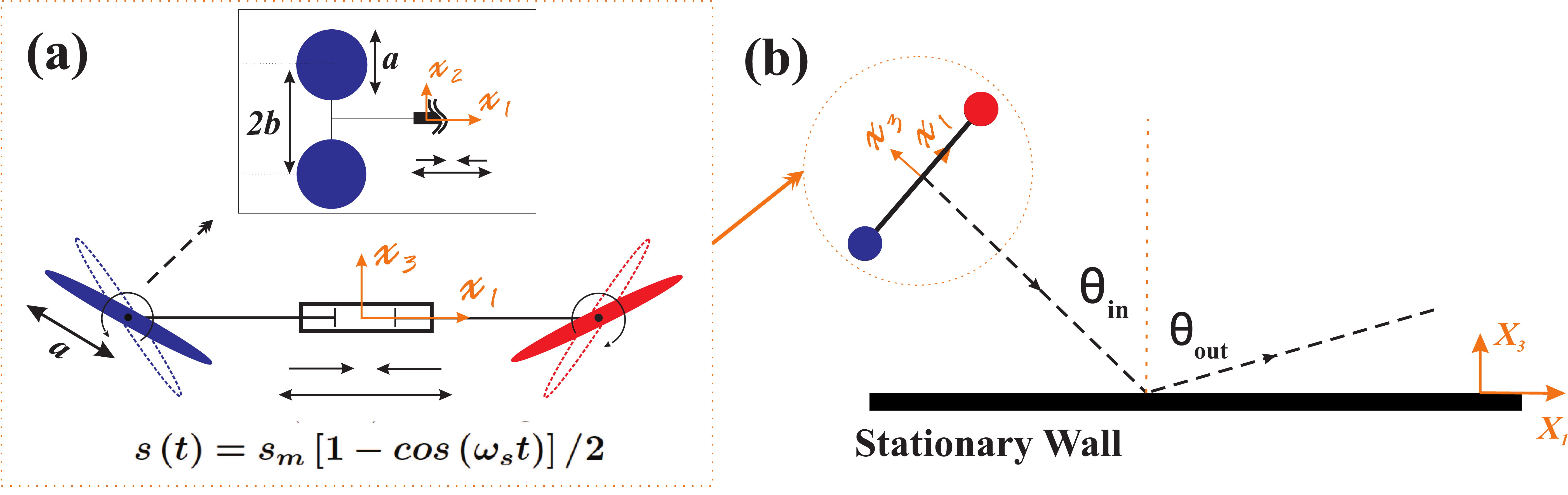}
	\\
	\includegraphics[width=0.65\textwidth]{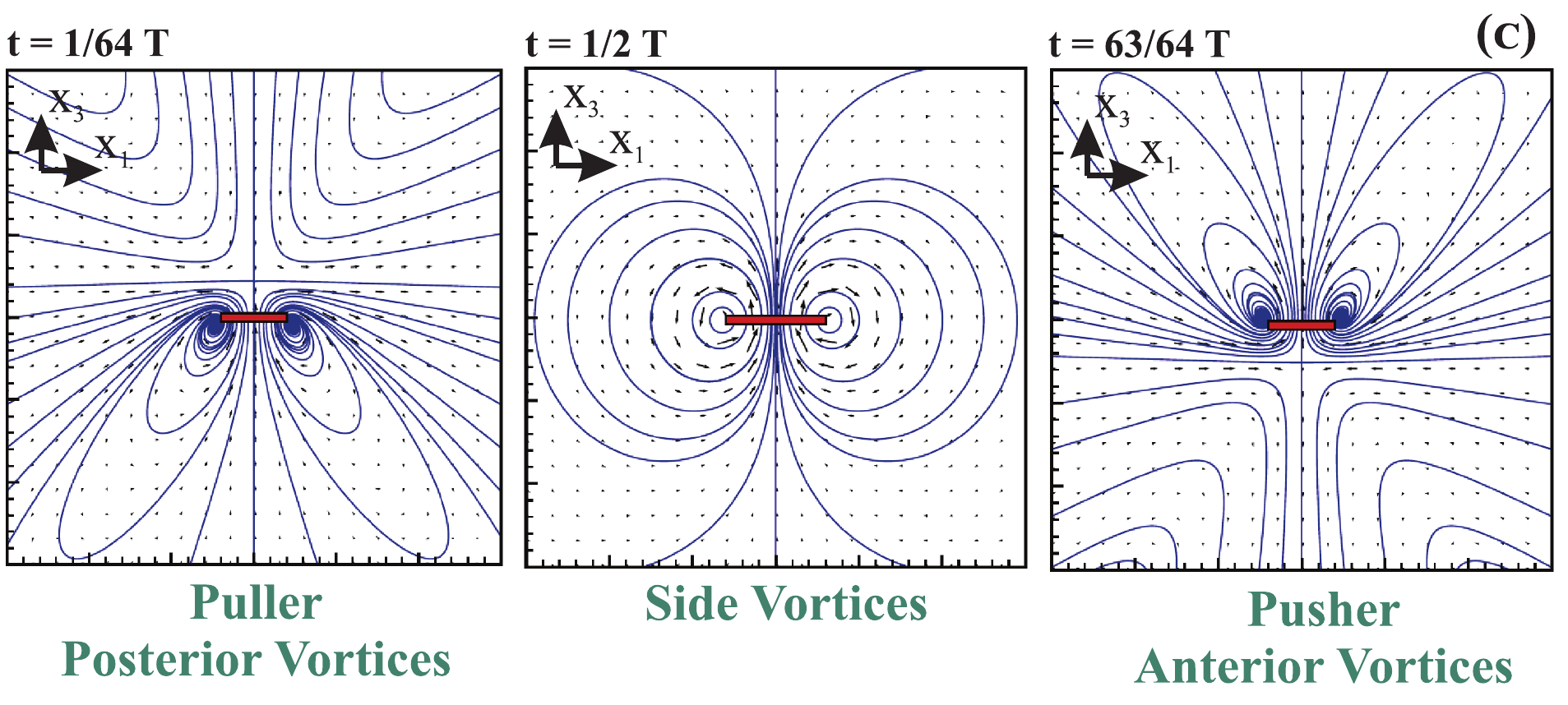}
	\caption {(a) Schematic of the model swimmer, which combines harmonic oscillation of its body length with counter-rotation of propellers. (b) Schematic representation of the model swimmer scattering off a stationary solid wall. $\theta_{in}$ and $\theta_{out}$ are defined with respect to the axis normal to the wall. (c) Snapshots of the oscillatory flow field induced by a single model swimmer in an infinite fluid \cite{jalali2015microswimmer}, which mimics the flow field around a C. reinhardtii cell \cite{guasto2010oscillatory}. The red thick bar represents chassis of the swimmer, blue lines demonstrate streamlines, and the time scale is $T=2\pi/\omega_s$.}
	\label{fig1}
\end{figure}

To mimic this complex flow field, a model microswimmer called \textit{Quadroar} has recently been proposed \cite{jalali2014versatile,jalali2015microswimmer,mirzakhanloo2018hydrodynamic}. The swimmer consists of two pairs of counter-rotating disks whose distance is periodically varied (Fig. \ref{fig1}-a). When all motions (reciprocating and rotating) have the same frequency and there is no phase difference, the model swimmer moves along a straight line in the $x_3$ direction, and induces an oscillatory flow field with  side, anterior and posterior vortices (Fig. \ref{fig1}-c). This flow field closely resembles the flow field of a C. reinhardtii cell \cite{jalali2015microswimmer,guasto2010oscillatory}. Specifically, oscillation of the linear actuator creates the oscillatory flow field between puller and pusher types, and the counter-rotation of disks contributes to the emergence of anterior, posterior, and side vortices. Via varying the relative frequency of propellers, or by imposing phase-differences between them, a full three dimensional reorientation maneuvers and tumblings can be obtained \cite{jalali2014versatile,jalali2015microswimmer}.

Let us consider a single swimmer moving near a no-slip solid boundary. The global frame of reference is fixed to the wall such that its $X_3$-axis is normal to the wall and points toward the semi-infinite fluid (Fig. \ref{fig1}-b). The swimmer's local frame of reference is attached to its geometric center so that its frame lies in $(x_1,x_2)$-plane, and $x_1$-axis is along the reciprocating chassis (Fig. \ref{fig1}-a). In our modelings, the length of each disk axle is denoted by $2b$, and reciprocating chassis' length is $2l+2s \lp t\rp$ where $s\lp t \rp = s_m \lb 1-cos\lp \omega_s t \rp \rb/2$, in which $s_m$ is the amplitude and $\omega_s$ is the frequency of oscillations. Angular velocities of the disks on left and right axles are $c_0 \omega_s$ and $- c_0 \omega_s$, where $c_0$ is a constant. We choose $b/a=l/a=4$ and $s_m/a=2$, and by choosing $\omega_s=1$, all frequencies in the problem are normalized by $\omega_s$. Here, unless otherwise noted, $c_0=50$ which is reminiscent of flagella beat for a C. reinhardtii cell ($\sim 50$ $Hz$). 

\begin{figure}[h]
	\centering
	\includegraphics[width=0.7\textwidth]{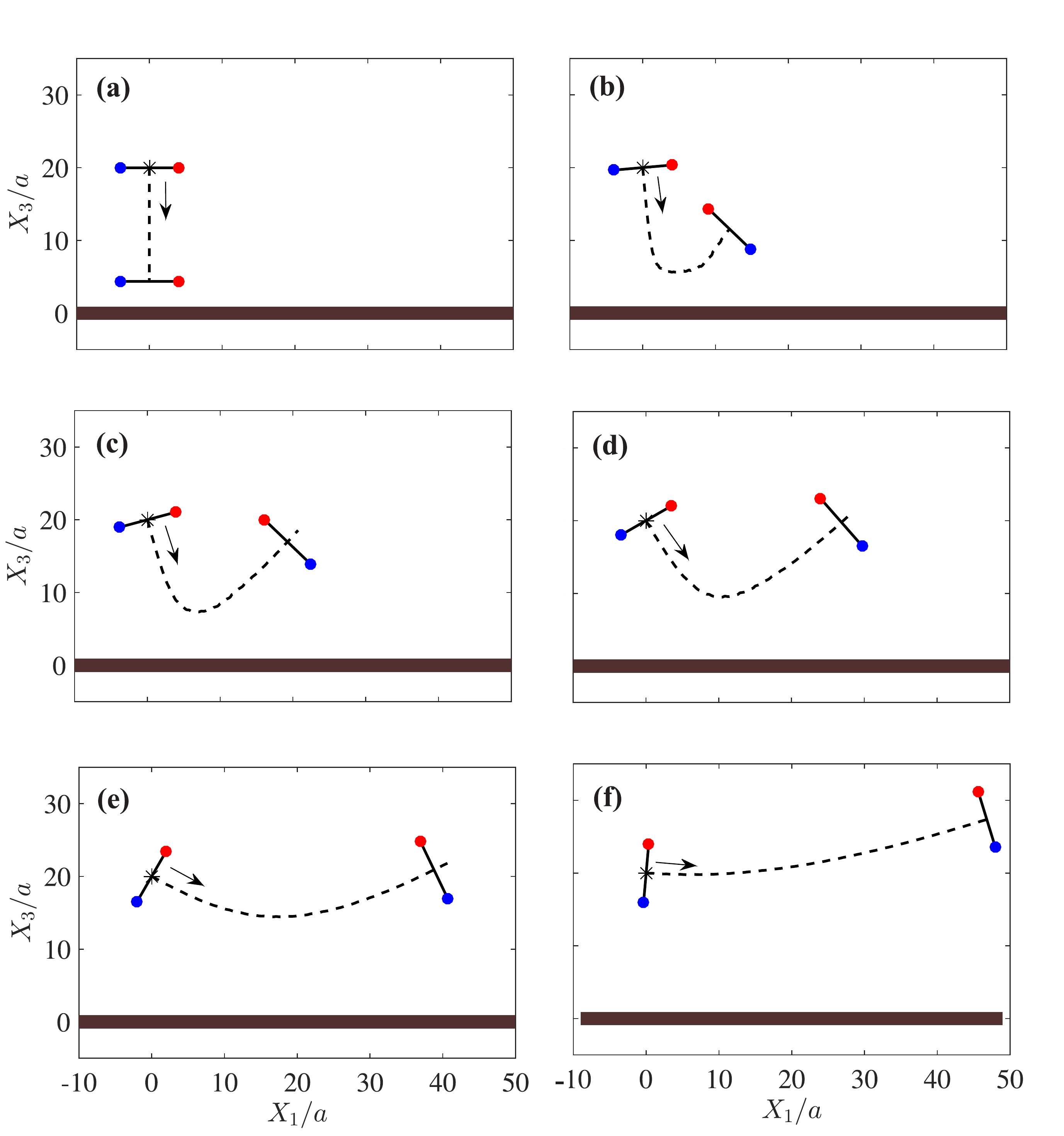}
	\caption
	{Samples of the hydrodynamic sensing and escaping behavior of microswimmers swimming near a solid boundary (denoted by the thick brown solid line at $X_3=0$). The swimmers initially swim toward the wall with different incidence angles: $\theta_{in}$ $=$ $0^o$ (a), $5^o$ (b), $15^o$ (c), $30^o$ (d), $60^o$ (e), and $85^o$ (f). The initial and final (after scattering) states of each case are shown. In each panel, the black thick bar represents the swimmer's body (c.f. Fig. \ref{fig1}-a), trajectory of the swimmer is shown by a dashed line, the start points are denoted by asterisks, and arrows represent the initial direction.} 
	\label{fig2}
\end{figure}

Contribution of each propeller (i.e. disk) to background streaming is modeled as the combination of point-force ($\fvec$) and point-torque ($\gammav$) flow fields. Therefore, our modeling involves four pairs of singularities (Stokeslets and rotlets) in the vicinity of a no-slip solid boundary. To satisfy the no penetration and no slip boundary condition on the wall, specific arrangements of singularities \cite{blake1971c,Blake1974} are placed at the image location (with respect to the solid wall) of each of the swimmer's singularities. It then can be shown (see appendix \ref{AppA}) that the velocity field due to a point-force near a no-slip wall is:
\begin{equation}\label{M2.11}
	\begin{multlined}
		u_i^f = \frac{f_j}{8\pi\eta} \lb \lp \frac{\delta_{ij}}{r}+\frac{r_i r_j}{r^3} \rp  - \lp \frac{\delta_{ij}}{\bar{r}}+\frac{\bar{r}_i \bar{r}_j}{\bar{r}^3} \rp \rb  +  \frac{2h f_j}{8\pi\eta} \lp \delta_{jm} \delta_{mk} - \delta_{j3}\delta_{3k}\rp \frac{\p}{\p \bar{r}_k} \lb \frac{h\bar{r}_i}{\bar{r}^3}-\lp \frac{\delta_{i3}}{\bar{r}} + \frac{\bar{r}_i \bar{r}_3}{\bar{r}^3} \rp \rb ,
	\end{multlined}
\end{equation}
where $\eta$ is dynamic viscosity and $\delta_{ij}$ is Kronecker delta. The point-force $\fvec$ is exerted at $\xvec_0=\lp \xi,\zeta,h \rp$, and the image point of $\xvec_0$ with respect to the stationary wall is given by $\bar{\xvec}_0 = \xvec_0 - 2 \lp \xvec_0 \cdot \bm{e}_3 \rp \bm{e}_3$, where $\bm{e}_3$ is the unit vector normal to the wall. Position of a generic point in space is denoted by vector $\xvec$, and $\rvec$ is defined as $\rvec= \xvec - \xvec_0$. Similarly, relative position of a generic point $\xvec$ from the image point $\bar{\xvec}_0$ is defined as $\bar{\rvec} = \xvec - \bar{\xvec}_0$. Using the same approach, velocity field due to a point-torque in the vicinity of a no-slip wall is derived as (see appendix \ref{AppA}):
\begin{equation}\label{M2.12}
	\begin{multlined}
		u_i^{\gamma} = \frac{1}{8\pi \eta} \lb \frac{\lp \bm{\gamma} \times \rvec \rp_i}{r^3} - \frac{\lp \bm{\gamma} \times \bar{\rvec} \rp_i}{\bar{r}^3} \rb   + \frac{1}{8\pi \eta} \lb 2h\epsilon_{kj3}\gamma_j \lp \frac{\delta_{ik}}{\bar{r}^3} - \frac{3 \bar{r}_i \bar{r}_k}{\bar{r}^5} \rp + 6\epsilon_{kj3} \frac{\gamma_j \bar{r}_i \bar{r}_k \bar{r}_3}{\bar{r}^5} \rb,
	\end{multlined}
\end{equation}
where the point-torque $\bm{\gamma}$ is exerted at $\xvec_0$ and all other parameters are defined in the same way as \eqref{M2.11}. Note that the velocity (vorticity) field at the position of each propeller, which in turn determines $\fvec$ or $\gammav$, is then the sum of contributions from all other propellers:
\begin{equation} \label{M2.21}
	\begin{multlined}
		\uvec_n = \sum_{k=1, k\neq n}^{4} \lp \uvec^f_k + \uvec^{\gamma}_k \rp, \quad 2\Omegav_n = \nabla \times \uvec_n,
	\end{multlined}
\end{equation}
where $\nabla \times$ is the curl operator, and $2\Omegav_n$ is the vorticity field at the center of propeller $n$. The force-free ($\sum_{k=1}^{4} \fvec_k = 0$) and torque-free ($\sum_{k=1}^{4} \lb \rvec_k \times \fvec_k + \gammav_k \rb= 0$) conditions in low-Reynolds-number regime, combined with velocity and vorticity fields presented in \ref{M2.21}, provide us with a closed system of thirty coupled equations and thirty unknowns that must be solved at each time step. Integrating linear and angular velocities in time will then provide the swimmer's position and orientation as a function of time (see appendix \ref{AppA}).

In our numerical experiments, the model swimmer is launched toward the wall with various incident angles $\theta_{in}$ (c.f. Fig.\ref{fig1}b). Scattering angle $\theta_{out}$ corresponding to each $\theta_{in}$ is then measured with respect to the normal to solid boundary after steady state is reached. We show samples of behavior of the swimmer for $\theta_{in}$ $=$ $0^o$, $5^o$, $15^o$, $30^o$, $60^o$, and $85^o$ in figure \ref{fig2}, in which the trajectory of the swimmer is shown by a black dashed line, chassis of the swimmer is denoted by a black thick bar, and the blue (red) filled circles represents propellers initially on the left (right) side of the swimmer.
Without even touching the wall, the swimmer feels the solid wall in all cases, and escapes the boundary similar to what has been observed experimentally for a C. reinhardtii cell \cite{kantsler2013ciliary,contino2015microalgae}. Note that sensing and escaping the boundary here is purely hydrodynamic, as there is no contact/flagellar force defined for the model swimmer.

The only exception in which the swimmer feels the boundary but cannot escape it, happens when a swimmer approaches the wall with $\theta_{in}=0$ (i.e. exactly normal to the wall). As theoretically required by the symmetry of our ideal numerical experiment, for $\theta_{in}=0$ the swimmer can not choose any direction over the other one. For a typical puller-type swimmer, far-field analysis predicts a head-on collision with the wall for this situation. But, here the swimmer has a complex oscillatory flow field in its close vicinity, which saves it from hitting the wall. Surprisingly, the swimmer stops swimming forward after getting close enough to the boundary (Fig. \ref{fig2}a). This state is, in fact, a \textit{dynamic} equilibrium. Because the swimmer is still struggling to swim forward with exactly the same stroke cycle as before and energy is getting wasted continuously through the propellers, but the time-averaged position of its geometric center has come to a halt . 
\begin{figure}[h]
	\centering
	\includegraphics[width=0.65\textwidth]{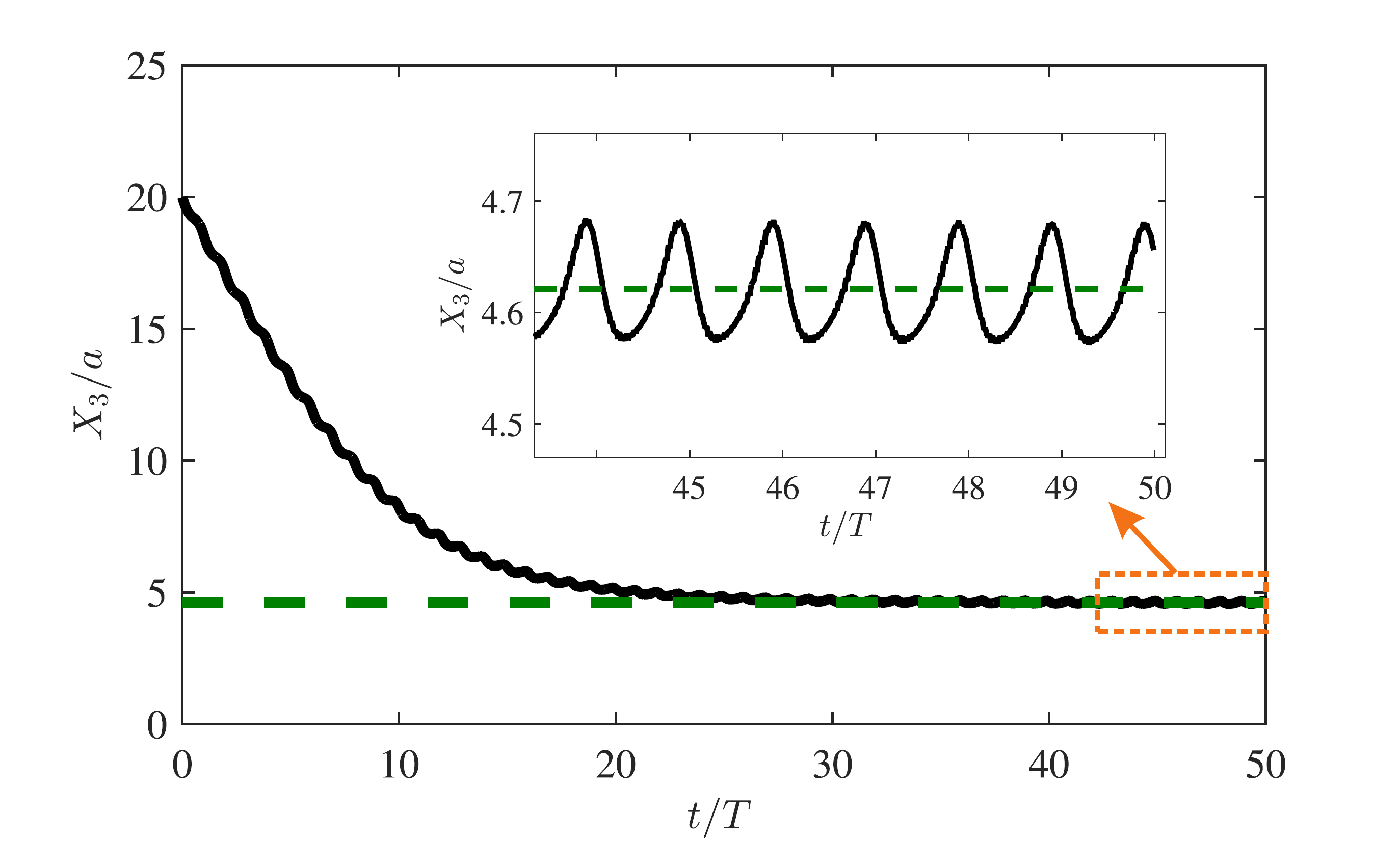}
	\caption
	{Time variation of the vertical position of a model swimmer approaching the wall with $\theta_{in}=0$, i.e. exactly normal to the wall. Inset is the zoomed view of the tail, which represents the very small-amplitude up-and-down oscillations.} 
	\label{fig3}
\end{figure}
Note that on very short length scales, there is an intrinsic oscillation in the trajectory of the model swimmer that originates from the oscillatory nature of its flow field. These small-amplitude ($\Delta Z/a \approx 0.1$) up-and-down oscillations (also reported for swimming C. reinhardtii cells as the `\textit{zigzagging motion}' \cite{garcia2011random}) will still be present in the dynamic equilibrium phase (see the inset of figure \ref{fig3}). However, there will be no net translation over time for the swimmer in this phase (see Fig. \ref{fig3}).

\begin{figure}[h]
	\centering
	\includegraphics[width=0.55\textwidth]{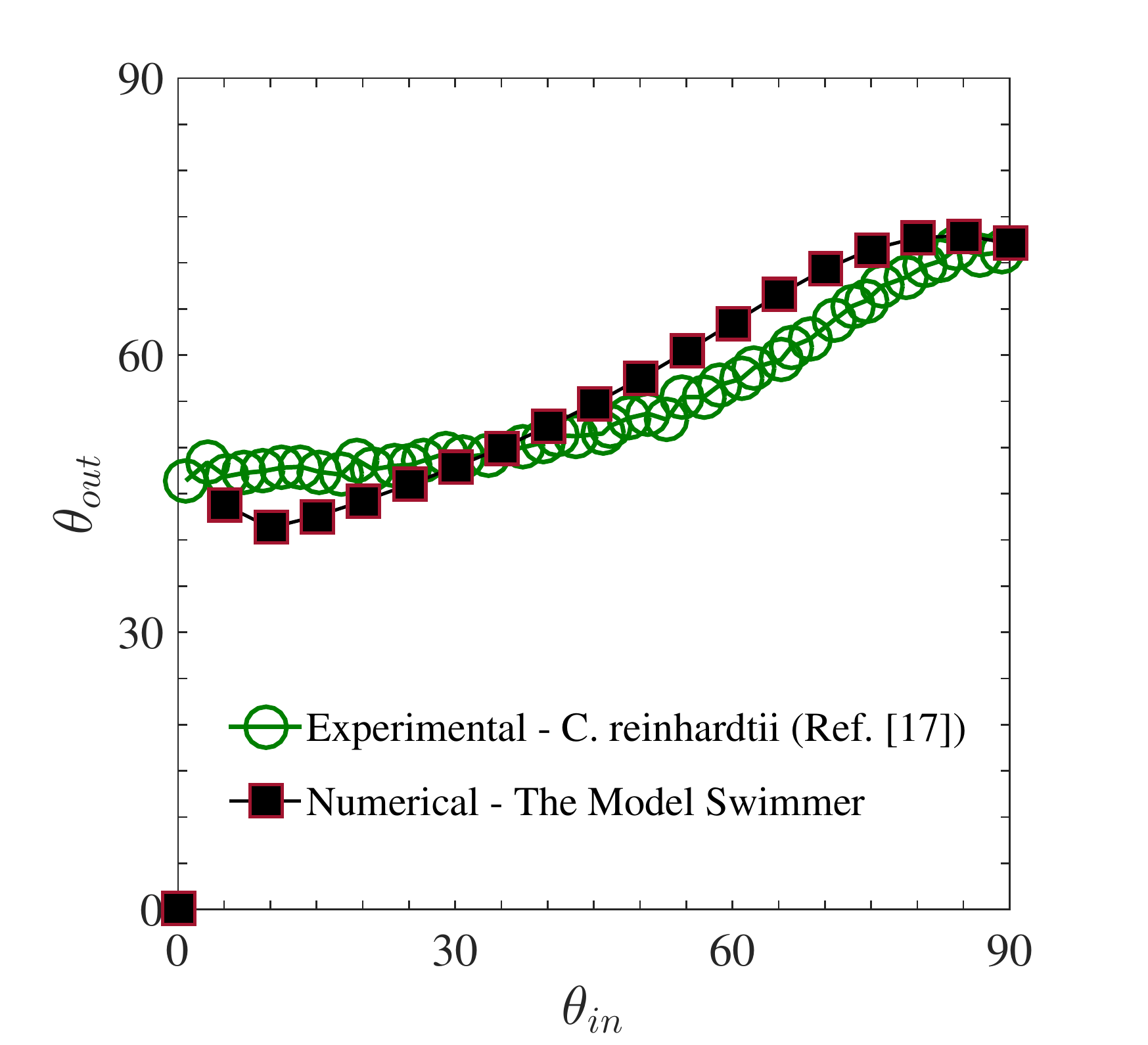}
	\caption
	{Comparison between the scattering angles ($\theta_{out}$) resulted from purely hydrodynamic numerical simulations of the model swimmer (black filled squares), and the experimental data (green circles) measured by \cite{contino2015microalgae} for wild type C. reinhardtii cells. Angles are presented in degrees.} 
	\label{fig4}
\end{figure}

Hydrodynamic scattering of our model swimmer, presented in the space of $\theta_{out}$ vs $\theta_{in}$, is in a very good agreement with a recent set of experimental data \cite{contino2015microalgae} on scattering of a real  wild-type C. reinhardtii cell (see figure \ref{fig4}). The only expected exception is at $\theta_{in}=0$ for which a perfect normal incidence (numerically easily achievable) results in a dynamical equilibrium, whereas such equilibrium has not been reported in the experiments, clearly due to extremely low probability of actual microorganisms approach the wall at the exact zero angle. 

In this letter, we demonstrated how inducing a complex oscillatory flow field (with anterior, side, and posterior vortices) is a sufficient tool for swimming cells to sense and escape the boundary. This clearly points to the hydrodynamic nature of surface-scattering. Our results are also in a very good agreement with recently released experimental data \cite{contino2015microalgae}. Our findings provide a new insight into the cell-surface scattering process. Also, may pave the path for new techniques in controlling biological migration, for which many potential applications (including diagnostics \cite{denissenko2012human}, drug delivery \cite{weibel2005microoxen}, and bioremediation \cite{valentine2010propane}) can be sought .

\newpage
\appendix

\section{Detailed Mathematical formulation } \label{AppA}

\subsection{Model swimmer in an infinite fluid: singularity solution} \label{sec2A} \hfill\\

Due to the micro-scale size of the swimmer, the corresponding Reynolds number is very small (i.e. $Re \ll 1$). Therefore, the effect of inertia is negligibly small compared to viscous effects, and Navier-Stokes equation of motion can be simplified to the Stokes equation:
\begin{equation} \label{2.23}
\nabla P = \eta \ \lap \uvec + \bm{F}, \quad \nabla \cdot \uvec = 0,
\end{equation}
where $P$ is the pressure filed, $\uvec$ is the velocity field, $\eta$ is dynamic viscosity of the ambient fluid, and $\bm{F}$ is the body force per unit volume. 

The model swimmer has four propellers (disks of radii $a$) which are placed at the ends of its left and right axles. Contribution of each disk to background streaming can be modeled as a combination of point-force ($\fvec$) and point-torque ($\gammav$) flow fields. The force and torque acting on each disk $i$ is given by 
\ba 
& \bm{f}_{i} = \eta  \ \mathscr{K}_{i} \cdot \lp \bm{v}_{i}-\bm{u}_{i}\rp, \label{eq2.1}\\
& \bm{\gamma}_{i} =\eta \ \mathscr{G} \cdot \lp \bm{\omega}_{i}-\bm{\Omega}_{i}\rp, \label{eq2.2}
\ea
where $\vvec_i$ and $\omegav_i$ are absolute linear and angular velocities of disk $i$; $\uvec_i$ and $2\Omegav_i$ are velocity and vorticity fields of the background fluid at the position of disk $i$, and $\eta$ is dynamic viscosity. The geometry of disks are hidden in $\mathscr{K}_i$,$\mathscr{G}$, which are tensors of rank two. Specifically, $\mathscr{K}_i$ is the translation tensor corresponding to disk $i$, and $\mathscr{G}$ is isotropic rotational tensor of a circular disk rotating about its diameter, with the forms given by \cite{Happel2012}:
\begin{eqnarray} \label{2.3}
 \mathscr{K}_{i}= 
\frac{8}{3} a \begin{bmatrix} 5-cos\lp 2 \alpha_{i}\rp & 0 & sin\lp 2 \alpha_{i}\rp\\ 0 & 4 & 0 \\ sin\lp 2 \alpha_{i}\rp & 0 & 5+cos\lp 2 \alpha_{i}\rp\end{bmatrix}
, \quad \mathscr{G}=\frac{32}{3}a^3 \bm{I},
\end{eqnarray}
where $\bm{I}$ is the identity tensor, $a$ is radius of each disk, and $\alpha_i$ denotes the angle that disk $i$ makes with $(x_1,x_2)$-plane of the swimmer. Considering only the point-force contribution of each propeller in an infinite fluid domain, the governing equation can be written as:
\begin{equation} \label{2.6}
\nabla P = \eta \ \lap \uvec + \fvec \delta (\rvec), \quad \nabla \cdot \uvec = 0,
\end{equation}
where $\delta \lp\rvec\rp$ is Dirac delta function. The point-force is exerted at $\xvec_0$, and for a generic point $\xvec$ in space $\rvec = \xvec-\xvec_0$ with $r= |\rvec|$. Equation \eqref{2.6} can be analytically solved in several ways (see e.g. \cite{chwang1975hydromechanics}), and the resultant velocity field is known as Stokeslet:  
\begin{equation} \label{2.16}
\uvec \lp \rvec,t \rp= \frac{\fvec}{8\pi \eta} \cdot \lp \frac{\bm{I}}{r}+\frac{\rvec \rvec}{r^3}\rp.
\end{equation}
The contribution of a point-torque $\bm{\gamma}$ exerted at a point $\xvec_0$ in an infinite fluid, on the other hand, is derived from the following set of equations \cite{chwang1975hydromechanics}: 
\begin{equation} \label{2.7}
\nabla P = \eta \ \lap \uvec + \nabla \times \lp \bm{\gamma} \delta (\rvec) \rp ,\quad \nabla \cdot \uvec = 0.
\end{equation}
The exact solution to \eqref{2.7} is also available (see e.g. \cite{chwang1975hydromechanics}), and is called a rotlet: 
\begin{equation} \label{2.17}
\uvec \lp \rvec,t \rp= \frac{1}{8\pi \eta}  \lp \frac{\gammav \times \rvec}{r^3} \rp.
\end{equation}
Linearity of Stokes equation allows us to invoke the principle of superposition. As a result, the net contribution of each disk (when placed in an unbounded fluid domain) to background streaming, can be modeled as the combination of a Stokeslet and a rotlet:
\begin{equation} \label{2.8}
\uvec \lp \rvec,t \rp= \frac{\fvec}{8\pi \eta} \cdot \lp \frac{\bm{I}}{r}+\frac{\rvec \rvec}{r^3} \rp + \frac{1}{8\pi \eta} \lp \frac{\gammav \times \rvec}{r^3} \rp.
\end{equation}
The velocity field that the model swimmer induces in its surrounding (when swimming in an infinite fluid domain) is then the sum of contributions from all of its disks:
\begin{equation} \label{2.4}
\uvec \lp \xvec,t \rp= \frac{1}{8\pi \eta} \sum_{k=1}^{4} \lp \frac{\fvec_k}{r_k}+\frac{\fvec_k \cdot \rvec_k}{r_k^3}\rvec_k+\frac{\gammav_k \times \rvec_k}{r_k^3} \rp,
\end{equation}
where $\xvec$ is the position vector of a generic point in space, and $\rvec_k$ is the vector connecting geometric center of disk $k$ to this point. To calculate the induced vorticity field, one needs to then take curl of the velocity field ($2\Omegav = \nabla \times \uvec$):
\begin{equation} \label{2.5}
2\Omegav \lp \xvec,t \rp = %\nabla \times \uvec =  
\frac{1}{8\pi \eta} \sum_{k=1}^{4} \lb \frac{2\fvec_k \times \rvec_k}{r_k^3}+\frac{3(\gammav_k \cdot \rvec_k) \rvec_k -r_k^2\gammav_k}{r_k^5} \rb.
\end{equation}

\subsection{Model swimmer in vicinity of a no-slip solid boundary} \hfill\\

As discussed in previous section, contribution of each disk to the background streaming is modeled here as a combination of a point-force and a point-torque. Therefore, our model swimmer involves four pairs of singularities. In the vicinity of a no-slip solid boundary, to satisfy the no-penetration and no-slip boundary conditions on the wall, a specific arrangement of singularities -- called image systems \cite{blake1971c,Blake1974} -- is placed at each singularity's image location.
The image systems of a Stokeslet ($\fvec$) that is parallel to and at a distance $h$ from a wall is a combination of a Stokeslet ($-\fvec$), Stokes-doublet ($2h\fvec$), and a source-doublet ($-4\eta h^2 \fvec$). For a rotlet($\bm{\gamma}$) at a distance $h$ and parallel to a wall the image system includes a rotlet ($-\bm{\gamma}$), a stresslet ($16\pi \eta \bm{\gamma}$), and a source-doublet ($8 \pi h \bm{\gamma}$). For a rotlet ($\bm{\gamma}$) that is normal to a wall the image system is just a single rotlet ($-\bm{\gamma}$), but for a Stokeslet ($\fvec$)  normal to a wall the image system includes a Stokeslet ($-\fvec$), a Stokes-doublet ($-2h\fvec$), and a source-doublet ($4\eta h^2 \fvec$).
Stokes-doublet, characterized by a strength tensor of rank two ($D_{jk}$), is (see e.g. \cite{Blake1974,batchelor1970stress}):
\begin{equation} \label{2.9}
u_i = \frac{D_{jk}}{8\pi \eta} \lb \lp -\frac{r_i \delta_{jk}}{r^3}+\frac{3 r_i r_j r_k}{r^5} \rp + \lp \frac{r_k\delta_{ij}-r_j\delta_{ik}}{r^3} \rp \rb.
\end{equation}
In the case of a force dipole which is symmetric and contributes no net torque to the surrounding fluid, the solution is called stresslet and can simply be defined as the symmetric part of a Stokes-doublet (first term on the right-hand-side of equation \eqref{2.9}):
\begin{equation} \label{2.26}
u_i^{sym} = \frac{D_{jk}}{8\pi \eta} \lp -\frac{r_i \delta_{jk}}{r^3}+\frac{3 r_i r_j r_k}{r^5} \rp.
\end{equation}
On the other hand, the skew-symmetric part of a Stokes-doublet \eqref{2.9} represents the net torque contribution of a force dipole. Thus, it is equivalent to the rotlet solution:  
\begin{equation} \label{2.10}
\nonumber u_i^{skew} = \frac{D_{jk}}{8\pi \eta} \lp \frac{r_k\delta_{ij}-r_j\delta_{ik}}{r^3} \rp \equiv \frac{1}{8\pi \eta} \frac{\lp \bm{\gamma} \times \rvec \rp _i}{r^3},
\end{equation}
where $\gamma_i = -\epsilon_{ijk} D_{jk}$.
Finally, the flow field due to a point-source with outward mass flux $M$ is $u_i=({M}/{4\pi})({r_i}/{r^3})$. Therefore, the velocity field due to a source-doublet can be written as: 
\begin{equation} \label{2.13}
u_i = \frac{M_{j}}{4\pi} \lp -\frac{\delta_{ij}}{r^3}+\frac{3 r_i r_j}{r^5} \rp .
\end{equation}

Using \eqref{2.16}, \eqref{2.9}, and \eqref{2.13} as the elements of our image system for a Stokeslet, velocity field due to a point-force near a stationary no-slip wall is obtained as \cite{blake1971c}: 
\begin{equation}\label{2.11}
\begin{multlined}
u_i^f = \frac{f_j}{8\pi\eta} \lb \lp \frac{\delta_{ij}}{r}+\frac{r_i r_j}{r^3} \rp  - \lp \frac{\delta_{ij}}{\bar{r}}+\frac{\bar{r}_i \bar{r}_j}{\bar{r}^3} \rp \rb  +  \frac{2h f_j}{8\pi\eta} \lp \delta_{jm} \delta_{mk} - \delta_{j3}\delta_{3k}\rp \frac{\p}{\p \bar{r}_k} \lb \frac{h\bar{r}_i}{\bar{r}^3}-\lp \frac{\delta_{i3}}{\bar{r}} + \frac{\bar{r}_i \bar{r}_3}{\bar{r}^3} \rp \rb ,
\end{multlined}
\end{equation}
where $\eta$ is dynamic viscosity and $\delta_{ij}$ is Kronecker delta. The point-force $\fvec$ is exerted at $\xvec_0=\lp \xi,\zeta,h \rp$, and the image point of $\xvec_0$ with respect to the stationary wall is given by $\bar{\xvec}_0 = \xvec_0 - 2 \lp \xvec_0 \cdot \bm{e}_3 \rp \bm{e}_3$, where $\bm{e}_3$ is the unit vector normal to the wall. Position of a generic point in space is denoted by vector $\xvec$, and $\rvec= \xvec - \xvec_0$. Similarly, relative position of a generic point $\xvec$ from the image point $\bar{\xvec}_0$ is defined as $\bar{\rvec} = \xvec - \bar{\xvec}_0$.
Here $m \in \lcb 1,2 \rcb$, and the expression $\delta_{jm} \delta_{mk} - \delta_{j3}\delta_{3k}$ is non-zero only if $j=k$. Then it is equal to $-1$ if $j=k=3$, and equal to $+1$ if $j=k=1\ or \ j=k=2$. Equation \eqref{2.11} can be also written in the familiar form of $u_i^f = G_{ij} f_j$, where $G_{ij} \lp \rvec ,\bar{\rvec} \rp$ stands for the free space Green's function of the Stokes equation:
\ba\label{2.15B}
\nonumber  G_{ij} \lp \rvec,\bar{\rvec} \rp =  &\frac{1}{8\pi\eta} \lb \lp \frac{\delta_{ij}}{r}+\frac{r_i r_j}{r^3} \rp  - \lp \frac{\delta_{ij}}{\bar{r}}+\frac{\bar{r}_i \bar{r}_j}{\bar{r}^3} \rp \rb  
+ \frac{1}{8\pi\eta} \lb 2h^2 \lp 1-2\delta_{j3} \rp \lp \frac{\delta_{ij}}{\bar{r}^3} - \frac{3 \bar{r}_i \bar{r}_j}{\bar{r}^5} \rp \rb \\
&+ \frac{1}{8\pi\eta} \lb 2h \lp 1-2\delta_{j3} \rp \lp \frac{\bar{r}_j \delta_{i3}}{\bar{r}^3} + \frac{3 \bar{r}_i \bar{r}_j \bar{r}_3}{\bar{r}^5} - \frac{\bar{r}_3 \delta_{ij}}{\bar{r}^3} - \frac{\bar{r}_i \delta_{j3}}{\bar{r}^3} \rp \rb .
\ea
Similarly, upon substituting \eqref{2.17}, \eqref{2.26}, and \eqref{2.13} into the image system of a rotlet, the velocity field of a point-torque in the vicinity of a stationary no-slip wall is then derived as \cite{Blake1974}:
\begin{equation}\label{2.12}
\begin{multlined}
u_i^{\gamma} = \frac{1}{8\pi \eta} \lb \frac{\lp \bm{\gamma} \times \rvec \rp_i}{r^3} - \frac{\lp \bm{\gamma} \times \bar{\rvec} \rp_i}{\bar{r}^3} \rb  + \frac{1}{8\pi \eta} \lb 2h\epsilon_{kj3}\gamma_j \lp \frac{\delta_{ik}}{\bar{r}^3} - \frac{3 \bar{r}_i \bar{r}_k}{\bar{r}^5} \rp + 6\epsilon_{kj3} \frac{\gamma_j \bar{r}_i \bar{r}_k \bar{r}_3}{\bar{r}^5} \rb.
\end{multlined}
\end{equation}

To sum up, for our model swimmer when swimming in vicinity of a solid boundary, the contribution of each propeller (p) to background streaming is given by:
\bsa\label{2.19}
& \uvec_p \lp \rvec,\bar{\rvec},t \rp= \uvec^f_p \lp \rvec,\bar{\rvec},t  \rp + \uvec^{\gamma}_p \lp \rvec,\bar{\rvec},t  \rp, \\
& 2\Omegav_p \lp \rvec,\bar{\rvec},t \rp= \nabla \times \lb \uvec^f_p\lp \rvec,\bar{\rvec},t \rp + \uvec^{\gamma}_p \lp \rvec,\bar{\rvec},t \rp \rb,
\esa
where $\uvec^f$ and $\uvec^{\gamma}$ are given by \eqref{2.11} and \eqref{2.12}. Note that the velocity (vorticity) field at the position of propeller n, which in turn determines $\fvec_n$ or $\gammav_n$, is the sum of contributions from all other propellers:
\begin{equation} \label{2.21}
\begin{multlined}
\uvec_n = \sum_{k=1, k\neq n}^{4} \lp \uvec^f_k + \uvec^{\gamma}_k \rp, \quad 2\Omegav_n = \nabla \times \uvec_n,
\end{multlined}
\end{equation}
where $2\Omegav_n$ is the vorticity field at the center of disk $n$.

The force-free ($\sum_{k=1}^{4} \fvec_k = 0$) and torque-free ($\sum_{k=1}^{4} \lp \rvec_k \times \fvec_k + \gammav_k \rp= 0$) conditions in low-Reynolds-number regime, combined with velocity and vorticity fields presented in \eqref{2.21}, provide us with a closed system of thirty coupled equations and thirty unknowns that must be solved at each time step. Integrating linear and angular velocities in time, using RK78 method \cite{fehlberg1968classical}, will then provide the swimmer's position and orientation as a function of time.

\section{Exploring the space of parameters} \label{AppB}

\begin{figure}[h]
	\centering
	\includegraphics[width=0.55\textwidth]{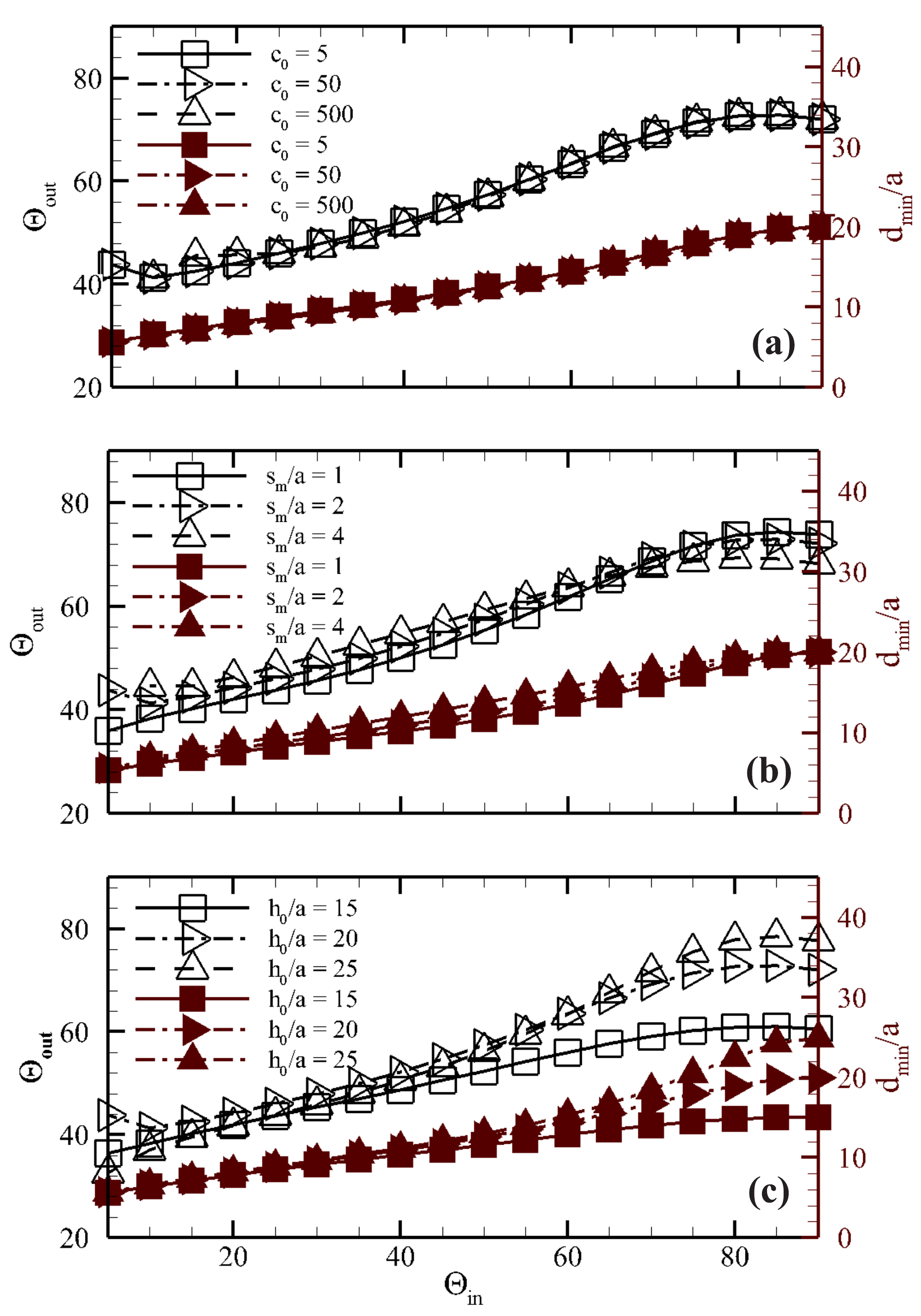}
	\caption
	{The scattering angle ($\theta_{out}$) of the model swimmer and its minimum distance ($d_{min}$) from the wall as a function of its incidence angle ($\theta_{in}$). The black (brown) axis on the left (right) measures the scattering angles (minimum distances from the wall). For each incidence angle, the scattering angle (minimum distance from the wall) of the swimmer is shown by a black unfilled (brown filled) square, right-triangle, and up-triangle for: (a) $c_0=$ 5, 50, and 500, respectively; (b) $s_m/a=$ 1, 2, and 4, respectively; and (c) $h_0/a=$ 15, 20, and 25, respectively. The benchmark (also presented in figure \ref{fig4}) corresponds to $c_0=50$, $s_m/a=2$, and $h_0/a=20$.} 
	\label{fig6}
\end{figure}

The numerical results presented in figure \ref{fig4} correspond to the case of $c_0=50$ which is reminiscent of the flagellar beat frequency of green alga C. reinhardtii. Our numerical experiments show that the scattering behavior of the model swimmer will remain the same for different values of $c_0$. Changing the value of $c_0$ will only change swimming speed of the swimmer, and thus the time required for its scattering. The swimmer's scattering angle ($\theta_{out}$) and its minimum distance ($d_{min}$) from the wall, as a function of its incidence angle ($\theta_{in}$), are presented in figure \ref{fig6}-a for different values of $c_0$. The results further confirm similarity of the behavior for swimmers with different propeller speeds. A similar effect has been also reported for amoeboid swimmers \cite{wu2015amoeboid,wu2016amoeboid}, where swimming stroke frequency does not change the navigation behavior.

Moreover, in our numerical experiments we have considered, as a benchmark, a model swimmer with $s_m/a=2$ so that $a=1\ \mu m$ provides us with the body size of $8-12 \ \mu m$ similar to that of a C. reinhardtii cell \cite{goldstein2015green}. To explore the effect of body size on the scattering behavior of the swimmer, figure \ref{fig6}-b represents the scattering angles ($\theta_{out}$) and minimum distances from the wall ($d_{min}$) of swimmers with different values of $s_m$. Qualitative and quantitative similarity of the scattering results observed for swimmers with different values of $s_m$, further highlights the primary role of flow characteristics (i.e. oscillatory nature of the flow combined with side, posterior, and anterior vortices) rather than body size of the swimmer.

Lastly, we also present the scattering results of the swimmer when launched toward the boundary with different initial distances ($h_0$) from the wall (figure \ref{fig6}-c). Note that by increasing the incidence angle ($\theta_{in}$), the effect of initial distance becomes more clear. In its extreme case, for $\theta_{in}=90^o$ (i.e. when the model swimmer initially swims parallel to the wall) the minimum distance is equal to $h_0$ itself, which here ie set to different values. As we get closer to the other extreme (i.e. swimming normal to the wall), quantitative difference between the results fades out both for the case of scattering angles and minimum distances from the wall (see figure \ref{fig6}-c).

\section*{Acknowledgments}
Authors acknowledge the support of National Science Foundation (NSF) via the grant No. CMMI-1562871.

%\bibliographystyle{unsrt}
%\bibliography{references} % Produces the bibliography via BibTeX.

\bibliographystyle{unsrt}
%\bibliography{HG,Bragg3,suntans,MyPapers}
%%%%%%%%%%%%%%%%%%%%%%%%%%%%%%%%%%%%%%%%%%%%%%%%%%%%%%%%%%%%%%%%%%%%%%%%%%%%%%%%%%%%
%%%%%%%%%%%%%%%%%%%%%%%%%%%%%%%%%%%%%%%%%%%%%%%%%%%%%%%%%%%%%%%%%%%%%%%%%%%%%%%%%%%%
%%%%%%%%%%%%%%%%%%%%%%%%%%%%%%%%%%%%%%%%%%%%%%%%%%%%%%%%%%%%%%%%%%%%%%%%%%%%%%%%%%%%
% \bibliography{Transformation_Optics.bib}

\begin{thebibliography}{10}
	
	\bibitem[Denissenko et~al., 2012]{denissenko2012human}
	Petr Denissenko, Vasily Kantsler, David~J Smith, and Jackson Kirkman-Brown.
	\newblock {Human spermatozoa migration in microchannels reveals
		boundary-following navigation.}
	\newblock {\em Proceedings of the National Academy of Sciences},
	109(21):8007--8010, 2012.
	
	\bibitem[Durham et~al., 2012]{durham2012division}
	William M. Durham, Olivier Tranzer, Alberto Leombruni, and Roman Stocker.
	\newblock {Division by fluid incision: Biofilm patch development in porous
		media.}
	\newblock {\em Physics of Fluids}, 24(9):091107, 2012.
	
	\bibitem[Berke et~al., 2008]{berke2008hydrodynamic}
	Allison~P Berke, Linda Turner, Howard~C Berg, and Eric Lauga.
	\newblock {Hydrodynamic attraction of swimming microorganisms by surfaces.}
	\newblock {\em Physical Review Letters}, 101(3):038102, 2008.
	
	\bibitem[Li et~al., 2009]{li2009accumulation}
	Guanglai Li and Jay X. Tang.
	\newblock {Accumulation of microswimmers near a surface mediated by collision
		and rotational brownian motion.}
	\newblock {\em Physical review letters}, 103(7):078101, 2009.
	
	\bibitem[Nash et~al., 2010]{nash2010run}
	R.W. Nash, R. Adhikari, J. Tailleur, and M.E. Cates.
	\newblock {Run-and-tumble particles with hydrodynamics: Sedimentation, trapping,
		and upstream swimming.}
	\newblock {\em Physical review letters}, 104(25):258101, 2010.
	
	\bibitem[Mino et~al., 2011]{mino2011enhanced}
	Gast{\'o}n Mi{\~n}o, Thomas~E Mallouk, Thierry Darnige, Mauricio Hoyos, Jeremi
	Dauchet, Jocelyn Dunstan, Rodrigo Soto, Yang Wang, Annie Rousselet, and Eric
	Clement.
	\newblock {Enhanced diffusion due to active swimmers at a solid surface.}
	\newblock {\em Physical review letters}, 106(4):048102, 2011.
	
	\bibitem[Drescher et~al., 2011]{drescher2011fluid}
	Knut Drescher, J{\"o}rn Dunkel, Luis~H Cisneros, Sujoy Ganguly, and Raymond E.
	Goldstein.
	\newblock {Fluid dynamics and noise in bacterial cell--cell and cell--surface
		scattering.}
	\newblock {\em Proceedings of the National Academy of Sciences},
	108(27):10940--10945, 2011.
	
	\bibitem[Pimponi et~al., 2016]{pimponi2016hydrodynamics}
	Pimponi, Daniela and Chinappi, Mauro and Gualtieri, Paolo and Casciola, Carlo Massimo.
	\newblock {Hydrodynamics of flagellated microswimmers near free-slip interfaces}
	\newblock {\em Journal of Fluid Mechanics}, 789:514--533, 2016.
	
	\bibitem[Guccione et~al., 2017]{guccione2017diffusivity}
	Guccione, Giorgia and Pimponi, Daniela and Gualtieri, Paolo and Chinappi, Mauro.
	\newblock {Diffusivity of E. coli-like microswimmers in confined geometries: The role of the tumbling rate.}
	\newblock {\em Physical Review E}, 96(4):042603, 2017.
	
	\bibitem[Molaei et~al., 2014]{molaei2014failed}
	Mehdi Molaei, Michael Barry, Roman Stocker, and Jian Sheng.
	\newblock {Failed escape: solid surfaces prevent tumbling of escherichia coli.}
	\newblock {\em Physical review letters}, 113(6):068103, 2014.
	
	\bibitem[Sipos et~al., 2015]{sipos2015hydrodynamic}
	Orsolya Sipos, K. Nagy, R. Di Leonardo, and P. Galajda.
	\newblock {Hydrodynamic trapping of swimming bacteria by convex walls.}
	\newblock {\em Physical review letters}, 114(25):258104, 2015.
	
	\bibitem[Wu et~al., 2015]{wu2015amoeboid}
	Wu, Hao and Thi{\'e}baud, Marine and Hu, W-F and Farutin, Alexander and Rafa{\"\i}, S and Lai, M-C and Peyla, Philippe and Misbah, Chaouqi
	\newblock {Amoeboid motion in confined geometry.}
	\newblock {\em Physical Review E}, 92(5):050701, 2015.
	
	\bibitem[Wu et~al., 2016]{wu2016amoeboid}
	Wu, Hao and Farutin, Alexander and Hu, Wei-Fan and Thi{\'e}baud, Marine and Rafa{\"\i}, Salima and Peyla, Philippe and Lai, Ming-Chih and Misbah, Chaouqi.
	\newblock {Amoeboid swimming in a channel.}
	\newblock {\em Soft matter}, 12(36):7470--7484, 2016.
	
	\bibitem[de et~al., 2016]{de2016understanding}
	de Graaf, Joost and Mathijssen, Arnold JTM and Fabritius, Marc and Menke, Henri and Holm, Christian and Shendruk, Tyler N.
	\newblock {Understanding the onset of oscillatory swimming in microchannels.}
	\newblock {\em Soft matter}, 12(21):4704--4708, 2016.	
	
	\bibitem[Kantsler et~al., 2013]{kantsler2013ciliary}
	Vasily Kantsler, J{\"o}rn Dunkel, Marco Polin, and Raymond E. Goldstein.
	\newblock {Ciliary contact interactions dominate surface scattering of swimming
		eukaryotes.}
	\newblock {\em Proceedings of the National Academy of Sciences},
	110(4):1187--1192, 2013.
	
	\bibitem[Lushi et~al., 2017]{lushi2017scattering}
	Enkeleida Lushi, Vasily Kantsler, and Raymond E. Goldstein.
	\newblock {Scattering of biflagellate microswimmers from surfaces.}
	\newblock {\em Physical Review E}, 96(2):023102, 2017.
	
	\bibitem[Contino et~al., 2015]{contino2015microalgae}
	Matteo Contino, Enkeleida Lushi, Idan Tuval, Vasily Kantsler, and Marco Polin.
	\newblock {Microalgae scatter off solid surfaces by hydrodynamic and contact
		forces.}
	\newblock {\em Physical review letters}, 115(25):258102, 2015.
	
	\bibitem[Jalali et~al., 2014]{jalali2014versatile}
	Mir~Abbas Jalali, Mohammad-Reza Alam, and SeyyedHossein Mousavi.
	\newblock Versatile low-reynolds-number swimmer with three-dimensional
	maneuverability.
	\newblock {\em Physical Review E}, 90(5):053006, 2014.
	
	\bibitem[Jalali et~al., 2015]{jalali2015microswimmer}
	Mir~Abbas Jalali, Atefeh Khoshnood, and Mohammad-Reza Alam.
	\newblock Microswimmer-induced chaotic mixing.
	\newblock {\em Journal of Fluid Mechanics}, 779:669--683, 2015.
	
	\bibitem[Guasto et~al., 2010]{guasto2010oscillatory}
	Jeffrey~S Guasto, Karl~A Johnson, and Jerry~P Gollub.
	\newblock Oscillatory flows induced by microorganisms swimming in two
	dimensions.
	\newblock {\em Physical review letters}, 105(16):168102, 2010.
	
	\bibitem[Drescher et~al., 2010]{drescher2010direct}
	Knut Drescher, Raymond~E Goldstein, Nicolas Michel, Marco Polin, and Idan
	Tuval.
	\newblock Direct measurement of the flow field around swimming microorganisms.
	\newblock {\em Physical Review Letters}, 105(16):168101, 2010.
	
	\bibitem[Mirzakhanloo et~al., 2018]{mirzakhanloo2018hydrodynamic}
	Mehdi Mirzakhanloo, Mir~Abbas Jalali, and Mohammad-Reza Alam.
	\newblock Hydrodynamic choreographies of microswimmers.
	\newblock {\em Scientific Reports}, 8(1):3670, 2018.
	
	\bibitem[Blake, 1971]{blake1971c}
	JR~Blake.
	\newblock A note on the image system for a stokeslet in a no-slip boundary.
	\newblock volume~70, pages 303--310. Cambridge University Press, 1971.
	
	\bibitem[Blake, 1974]{Blake1974}
	JR~Blake and AT~Chwang.
	\newblock Fundamental singularities of viscous flow.
	\newblock {\em Journal of Engineering Mathematics}, 8(1):23--29, 1974.
	
	\bibitem[Garcia et~al., 2011]{garcia2011random}
	Micha{\"e}l Garcia, Stefano Berti, Philippe Peyla, and Salima Rafa{\"\i}.
	\newblock Random walk of a swimmer in a low-reynolds-number medium.
	\newblock {\em Physical Review E}, 83(3):035301, 2011.
	
	\bibitem[Weibel et~al., 2005]{weibel2005microoxen}
	Douglas~B Weibel, Piotr Garstecki, Declan Ryan, Willow~R DiLuzio, Michael
	Mayer, Jennifer~E Seto, and George~M Whitesides.
	\newblock Microoxen: Microorganisms to move microscale loads.
	\newblock {\em Proceedings of the National Academy of Sciences},
	102(34):11963--11967, 2005.
	
	\bibitem[Valentine et~al., 2010]{valentine2010propane}
	David~L Valentine, John~D Kessler, Molly~C Redmond, Stephanie~D Mendes,
	Monica~B Heintz, Christopher Farwell, Lei Hu, Franklin~S Kinnaman, Shari
	Yvon-Lewis, Mengran Du, et~al.
	\newblock Propane respiration jump-starts microbial response to a deep oil
	spill.
	\newblock {\em Science}, 330(6001):208--211, 2010.
	
	\bibitem[Happel et~al., 2012]{Happel2012}
	John Happel and Howard Brenner.
	\newblock {\em Low Reynolds number hydrodynamics: with special applications to
		particulate media}, volume~1.
	\newblock Springer Science \& Business Media, 2012.
	
	\bibitem[Chwang et~al., 1975]{chwang1975hydromechanics}
	Allen~T Chwang and T~Yao-Tsu Wu.
	\newblock Hydromechanics of low-reynolds-number flow. part 2. singularity
	method for stokes flows.
	\newblock {\em Journal of Fluid Mechanics}, 67(4):787--815, 1975.
	
	\bibitem[Batchelor, 1970]{batchelor1970stress}
	GK~Batchelor.
	\newblock The stress system in a suspension of force-free particles.
	\newblock {\em Journal of fluid mechanics}, 41(3):545--570, 1970.
	
	\bibitem[Fehlberg, 1968]{fehlberg1968classical}
	Erwin Fehlberg.
	\newblock Classical fifth-, sixth-, seventh-, and eighth-order runge-kutta
	formulas with stepsize control.
	\newblock {\em National Aeronautics and Space Administration}, 1968.
	
	
	
	\bibitem[Goldstein, 2015]{goldstein2015green}
	Raymond E Goldstein
	\newblock Green algae as model organisms for biological fluid dynamics.
	\newblock {\em Annual review of fluid mechanics}, 47, 2015.

	
\end{thebibliography}

%%%%%%%%%%%%%%%%%%%%%%%%%%%%%%%%%%%%%%%%%%%%%%%%%%%%%%%%%%%%%%%%%%%%%%%%%%%%%%%%%%%%
%%%%%%%%%%%%%%%%%%%%%%%%%%%%%%%%%%%%%%%%%%%%%%%%%%%%%%%%%%%%%%%%%%%%%%%%%%%%%%%%%%%%
%%%%%%%%%%%%%%%%%%%%%%%%%%%%%%%%%%%%%%%%%%%%%%%%%%%%%%%%%%%%%%%%%%%%%%%%%%%%%%%%%%%%
\end{document}